\documentclass[twocolumn,pre,aps]{revtex4}
\usepackage{epsf}
\usepackage{graphicx}
\begin{document}
\title{Pattern forming instability induced by light in pure and
dye-doped nematic liquid crystals } 
\author{D.O. Krimer}
\author{G. Demeter}
\altaffiliation[]{on leave from the Research Institute for Particle
and Nuclear Physics of the Hungarian Academy of Sciences, Budapest,
Hungary.}
\author{L. Kramer}
\affiliation{Physikalisches Institut der Universit\"at Bayreuth, D-95440
Bayreuth,  Germany
}
\date{\today}
\begin{abstract}
We study theoretically the instabilities induced by a linearly polarized
ordinary light wave incident at a small oblique angle on a thin layer of
homeotropically oriented nematic liquid crystal with special emphasis on the
dye-doped case. The spatially periodic Hopf bifurcation that occurs as the
secondary instability after the stationary Freedericksz transition is analyzed.
\end{abstract}
\maketitle
\section{Introduction}
Liquid crystals (LCs) are known to demonstrate a very rich variety of
interesting optical phenomena which have been studied intensively
during the last two decades. A nematic LC behaves optically
as a uniaxial anisotropic  medium with the optical axis along the local
molecular orientation described by the director ${\bf n(r},t)$. Moreover, when
light propagates through the nematic, its electric field exerts a torque
on the molecules which can induce  molecular reorientation. When there is only
an ordinary wave in the LC (polarization perpendicular to the plane
containing the optical axis and wavevector), the initial distribution of the
director becomes unstable when the intensity of light reaches a certain
critical value. This is the so-called Light Induced Freedericksz  Transition
(LIFT). The director reorientation leads to a change of birefringence and, as
a consequence, the polarization is changed as light propagates through the
layer \cite{Tab85,simoni}.  

It is known \cite{mag_85,mag_87} that a periodic equilibrium
configuration of the nematic director can appear in a thin-film LC in the
magnetic or electric field induced Freedericksz transition under certain
conditions. Our work is devoted to the search of analogous phenomena in the
LIFT of nematic LCs including the dye-doped case. We will show that the Hopf
bifurcation that occurs as a secondary bifurcation after the LIFT leads indeed
to a periodic pattern, although the mechanism is here quite different (see
Conclusions). Doping is important because the LIFT threshold of a dye-doped
nematic can be two orders of magnitude smaller than for a pure nematic. The
nature of this anomalously low threshold was the subject of numerous studies
(see \cite{JAN_99,MAR_97}  and references therein). The fact that the threshold
intensity is  low allows the spot size of the light to be much larger than the
thickness of the layer, thus a large aspect ratio system can be realized. 

This paper is organized as follows. In Sec. \ref{theoret_model}
we present the theoretical description of our problem. In Sec.
\ref{section_lin} we perform the linear stability analysis of the homeotropic
state which gives the threshold for the LIFT. The numerical method of
calculating the stationary distorted state is described in Sec. \ref{st_dist}.
Finally, in Sec. \ref{stab_anal}, we do the linear stability analysis of the
stationary distorted  state with respect to general perturbations
in the plane of the nematic layer.
\section{Theoretical model}
\label{theoret_model}
We consider a linearly-polarized plane wave incident at a small
oblique angle $\beta_0$ on a layer of dye-doped nematic LC which has
initially homeotropic alignment (see Fig. 1). The light is polarized
along the $y$-axis i.e., we deal with an ordinary wave. Strong anchoring of
the nematic at the boundaries of the layer is assumed. The optical torque
acting on the director is given by  $\tau =\xi_{eff}/16\pi({\bf n}\cdot{\bf
E^{\ast}})  ({\bf n}\times{\bf E})+c.c.$, where ${\bf E}$ is the amplitude of
the optical electric field,  $\xi_{eff}=\varepsilon_a+\zeta$. 
$\varepsilon_a=\varepsilon_\|-\varepsilon_\perp$ is the dielectric
anisotropy and $\varepsilon_\perp {~}$($\varepsilon_\|$) is the dielectric
permittivity (at optical frequency) perpendicular (parallel) to ${\bf n}$.
$\zeta$ phenomenologically describes the effect of certain dye dopants
($\xi_{eff}=\varepsilon_a$ in a pure LC) and can be both positive and
negative and  depends on dye concentration, molecular structures of both host
and dye materials, on the  wavelength of light, and on the temperature 
\cite{JAN_99,MAR_97}. Obviously, the electrical part of the free energy will 
contain the same factor $\xi_{eff}$. The density of the free energy of the
dye-doped nematic LC is thus assumed in the form: 
\begin{eqnarray}
\label{free_energy}
F = F_{elastic}-\frac{\xi_{eff}}{16\pi}\mid {\bf n}\cdot{\bf E}
\mid^2, 
\end{eqnarray} 
where $F_{elastic}=K_1/2\left({\bf \nabla}\cdot{\bf n}\right)^2+
K_2/2\left({\bf n}\cdot{\bf \nabla}\times{\bf n}\right)^2+
K_3/2\left({\bf n}\times{\bf \nabla}\times{\bf n}\right)^2$ is the 
standard Frank free energy density and $K_1,K_2,K_3$ are respectively the
splay, twist and bend elastic constants of the LC \cite{Gen}.

We first assume that the director components depend only on
$z,t$. We introduce the angles $\theta(z,t)$
and $\varphi(z,t)$ (see Fig.1) so that 
${\bf n}=\left(\sin\theta,\cos\theta\sin\varphi,\cos\theta\cos\varphi\right)$.
Using the standard variational principle \cite{Tab85} and taking the
dissipation function in the form $R={\gamma\over 2}\dot {\bf n}^2$, where
$\gamma$ is an effective rotational viscosity, the equations of
motion for $\theta(z,t)$ and $\varphi(z,t)$ can be derived. 

In addition we need Maxwell's equations to determine the
electric field. These equations contain the complex 
dielectric tensor which depends on the director components:
\begin{figure}
\epsfxsize=7.true cm
\makebox[\columnwidth]{\epsffile{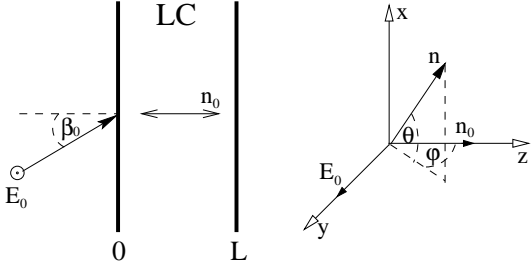}}
\caption{ Geometry of the setup: linearly polarized light along the ${\bf
y}$-direction incident at angle $\beta_0$ on a nematic LC layer with the
director ${\bf n_0} \parallel {\bf z} $ (homeotropic state). The components of
the director ${\bf n}$ are described in terms of the angles $\theta,\varphi$ 
($\theta=\varphi=0$ in the homeotropic state).}  
\end{figure}
\begin{eqnarray}
\label{tensor_e}
\varepsilon_{ij}=\left(\varepsilon_{\perp}+i\gamma_{\perp}\right) \delta_{ij}
+\left(\varepsilon_a+i\gamma_a \right) n_i n_j.
\end{eqnarray}
In (\ref{tensor_e}) $\gamma_a=\gamma_\perp-\gamma_\|$, where 
$\gamma_\perp$ and $\gamma_\|$ are the imaginary parts of the 
dielectric permittivity for ${\bf E}$ perpendicular and parallel to     
${\bf n}$ respectively. They describe the absorption effect by the dye, so
they vanish in pure LCs. The magnetic anisotropy at optical
frequencies can be neglected. Since the components of the dielectric tensor
depend on the $z$-coordinate only, we may use the stratified medium approach
for describing wave propagation \cite{simoni}. We write the electric and
magnetic fields in the form:  ${\bf E}({\bf r},t)=1/2({\bf
E}(z,t)e^{is_0k_0x}e^{-i\omega t}+c.c.), {\bf H}({\bf r},t)=1/2({\bf
H}(z,t)e^{is_0k_0x}e^{-i\omega t}+c.c.)$, where $k_0=\omega/c$ is the
wavenumber in vacuum and $s_0=\sin(\beta_0)$. Here ${\bf E}(z,t),{\bf H}(z,t)$
are amplitudes that vary slowly in time compared to $\omega^{-1}$ and obey the
equation:  
\begin{equation} 
\label{eq_psi}
{d\bar{\Psi}\over dz}=ik_0{\bf\sf D}{\bar\Psi},
\end{equation}
where 
\begin{equation}
\bar{\Psi}=\left(\matrix{E_x \cr H_y \cr E_y \cr -H_x}\right)
\end{equation}
and
\begin{equation}
{\bf\sf D}(z)=\left(\matrix{ 
 -{\varepsilon_{xz}s_0\over\varepsilon_{zz}} &
   1 - {\frac{{s_0^2}}
     {\varepsilon_{zz}}} & -{
      \frac{\varepsilon_{yz}\,s_0}
     {\varepsilon_{zz}}} & 0 \cr
   \varepsilon_{xx} - 
   {\frac{{{\varepsilon_{xz}}^2}}
     {\varepsilon_{zz}}} & -{
      \frac{\varepsilon_{xz}\,s_0}
     {\varepsilon_{zz}}} & 
    \varepsilon_{xy} - 
   {\frac{\varepsilon_{xz}\,
       \varepsilon_{yz}}{
       \varepsilon_{zz}}} & 0 \cr
   0 & 0 & 0 & 1 \cr \varepsilon_{xy} - {\frac{\varepsilon_{xz}\,\varepsilon_{yz}}{
       \varepsilon_{zz}}} & -
     {\frac{\varepsilon_{yz}\,s_0}
     {\varepsilon_{zz}}} & 
    \varepsilon_{yy} - 
   {\frac{{{\varepsilon_{yz}}^2}}
     {\varepsilon_{zz}}} - {s_0^2} &
   0 \cr  }\right).
\label{d}
\end{equation}
The $z$-component of the electric field can be found from
the following relation:
\begin{equation}
E_z=-{s_0\over\varepsilon_{zz}}H_y-{\varepsilon_{xz}
\over\varepsilon_{zz}}E_x
-{\varepsilon_{yz}\over\varepsilon_{zz}}E_y.
\label{ez}
\end{equation}
We will examine the case $\xi_{eff}>0$ so that the preferred orientation
corresponds to the director parallel to the electric field ${\bf n} || {\bf
E}$. Since in our geometry initially ${\bf n} \perp {\bf E}$, the homeotropic
state will cease to be stable above some critical intensity of the incident
light. The reorientation of the LC leads to modification of the electric
field polarization inside the LC owing to the fact that it becomes an
inhomogeneous anisotropic medium. 
\section{Stability analysis of the homeotropic state}
\label{section_lin}
We first perform the linear stability analysis of the homeotropic state 
($\theta=\varphi=0$). The linearized equation of motion for $\varphi(z,t)$  has
the following simple form:
\begin{eqnarray}
\label{phi}
&&\gamma \partial_t \varphi = K_3 \partial_z^2 \varphi
+\frac{(\varepsilon_a+\zeta)}{16\pi}
(2 \mid E_{0y} \mid^2\varphi + E_{1z}^{\ast}E_{0y}+
\nonumber\\
&&E_{1z}E_{0y}^{\ast}).
\end{eqnarray}
Here $E_{0y}$ is the $y$-component of the electric field amplitude for
the undistorted nematic (homeotropic orientation) and $E_{1z}$ is 
the $z$-component of the field that is caused by nematic 
reorientation (calculated to the first order in $\varphi$). It is easily
seen that in the undistorted LC the light maintains its
polarization inside the layer, so that we have only one nonzero component of
the electric field  $E_{0y}(z)= E_0 e^{ik_z z}$, where $k_z =k_{Re}+i k_{Im}
\simeq k_0  \sqrt{\varepsilon_{\perp}-s_0^2}+i\gamma_{\perp} k_0/ (2
\sqrt{\varepsilon_{\perp}-s_0^2})$ (terms of the order of 
$\left(\gamma_{\perp}/(\varepsilon_{\perp}-s_0^2)\right)^2$ in $k_z$
are neglected because $\gamma_{\perp}\ll\varepsilon_{\perp}$) and $E_0$ is the 
amplitude of the incident electric field. In the linear approximation
$\theta$ remains zero. Straightforward calculations yield the following
equation for $E_{1z}(z)$ from Eqs. (\ref{eq_psi}-\ref{ez}):
\begin{eqnarray}
\label{eq_E1z}
&&(\varepsilon_{\perp}+\varepsilon_a+i (\gamma_a+\gamma_{\perp}))
\frac{d^2 E_{1z}}{dz^2}+k_0^2 (\varepsilon_{\perp}+i \gamma_{\perp})
\nonumber\\
&&\times (\varepsilon_{\perp}+\varepsilon_a-s_0^2+i
(\gamma_a+\gamma_{\perp}))E_{1z}+ k_0^2 (\varepsilon_a+i \gamma_a)
\nonumber\\
&&\times (\varepsilon_{\perp}+i \gamma_{\perp})
\varphi E_{0y}+(\varepsilon_a+i \gamma_a)\frac{d^2( \varphi E_{0y})}{dz^2}
=0
\end{eqnarray}

Substituting $E_{1z}(z)$ into (\ref{eq_E1z}) in the form $E_{1z}(z)= E(z)
e^{ik_z z}$ and taking into account that $k_0 L \gg 1$ ($L$ is the width
of the layer), a first-order ODE  for $E(z)$ can be derived. Keeping in mind
that $E_{1z}(0)=0$ we eventually obtain from Eqs. (\ref{phi},\ref{eq_E1z}) 
the following  integro-differential equation for $\varphi$: 
\begin{eqnarray} 
\label{basic_1}
&&\tau\frac{\partial\varphi(z,t)}{\partial t}= \left(\frac{L}{\pi}\right)^2 
\frac{\partial^2\varphi(z,t)}{\partial z^2}+\rho\left\{ \left(\frac{\pi \kappa}{L}\right)
\right.
\nonumber\\
&& \left. 
\times \int_0^z \left[\psi \cos \left(\frac{\pi \kappa}{L} (z'-z)\right)+
\sin\left(\frac{\pi\kappa}{L} (z'-z)\right)\right]
\right.
\nonumber\\
&&\left.
\times e^{ \frac{\pi}{L} \xi \kappa (z'-z) } \varphi(z',t) dz' 
+\varphi(z,t) \right\} e^{-2 k_{Im} z },
\end{eqnarray}
where $\psi$, $\xi$, $\kappa$ and $\tau$ are parameters defined as:
\begin{eqnarray}
&&\psi = -\frac{\varepsilon_a^2 \gamma_{\perp} -
3 \varepsilon_a \varepsilon_{\perp} \gamma_{\perp}+
2 \gamma_a \varepsilon_{\perp}^2-
2 \gamma_ a \varepsilon_{\perp} \varepsilon_a }
{2 \left(\varepsilon_a+\varepsilon_{\perp} \right) 
\varepsilon_a \varepsilon_{\perp}} \nonumber{,~~}
\nonumber\\
&&\xi = \frac{2 \gamma_a \varepsilon_{\perp}^2 -
3 \varepsilon_a \varepsilon_{\perp} \gamma_{\perp}-
\varepsilon_a^2 \gamma_{\perp}}
{2 \left(\varepsilon_a+\varepsilon_{\perp} \right)
\varepsilon_a \varepsilon_{\perp}} 
\nonumber\\
&&\kappa={L\over\pi}{s_0^2\varepsilon_ak_0\over2\sqrt{\varepsilon_\perp}
(\varepsilon_\perp+\varepsilon_a)}\nonumber,{~}
\tau=\frac{\gamma L^2}{\pi^2 K_3}
\end{eqnarray}
In the parameters defined above only the 
linear terms in $\gamma_a,\gamma_{\perp}$ were kept ($\gamma_a,\gamma_{\perp} 
\ll 1$). The parameter $\tau$  is the characteristic time of the director
motion  and $\rho=I/I_c$, where $I$ is the intensity of the incident light and
$I_c$ is defined as:
\begin{eqnarray}
\label{porog}
I_c={\pi^2\over L^2}
{c (\varepsilon_\perp+\varepsilon_a)K_3\over 
\varepsilon_a\sqrt{\varepsilon_\perp} \eta},{~}\eta =
(\varepsilon_a+\zeta)/\varepsilon_a 
\end{eqnarray}
$I_c$ coincides with the threshold intensity of the LIFT for a pure nematic
($\eta=1$, $\gamma_{\perp}=\gamma_{\|}=0$) at perpendicular incidence
\cite{simoni}. Then Eq. (\ref{basic_1}) reduces to one obtained in
\cite{Tab_98}.

We use a two-mode expansion with respect to $z$ 
for the angle $\varphi$ with the boundary conditions $\varphi(z=0)=\varphi(z=L)=0$:
$\varphi(z,t)=A_1(t)\sin \left( \pi z/L\right)+
A_2(t) \sin \left(2 \pi z/L\right)$,
where $A_1$ and $A_2$ are time-dependent amplitudes. This is motivated by the
fact that the distorted state is asymmetric with respect to
the center of the layer because of absorption and the perturbation of the light
polarization inside the layer. Therefore we have to include at least one mode
that is symmetric and one mode that is antisymmetric with respect to the
center of the layer. After projecting Eq.(\ref{basic_1}) onto the trial
functions we have a system of two equations for the modes $A_1$ and $A_2$: 
\begin{eqnarray}
\tau \frac{dA_1}{dt}= {\cal L}_{11} A_{1}+{\cal L}_{12} A_{2},\
\tau \frac{dA_2}{dt}= {\cal L}_{21} A_{1}+{\cal L}_{22} A_{2},
\end{eqnarray}
where the elements of the matrix ${\cal L}_{ij}$ depend on material parameters 
and the control parameters $\rho$ and $\kappa$ (which is proportional to
$s_0^2$). We look for solutions proportional to $exp(\sigma t)$, where
$\sigma$ is the growth rate. The procedure  of deriving ${\cal
L}_{ij}$ is straightforward but the expressions for  these elements are too
long to be presented here. 

The stability diagram in the $(\kappa,\rho)$ plane can now be calculated for
any given material parameters of the LC.  As an example we consider the
nematic 5CB doped with the dye AD1 at $0.1\%$ concentration. We used the
following values of material parameters at the temperature $T=24^{\circ}$: 
$\alpha_o=42\ cm^{-1}$, $n_o=1.53$, $\alpha_e=190\ cm^{-1}$, $n_e=1.71$,      
(absorption coefficients and refractive indices of the ordinary and
extraordinary light, respectively),  $\lambda = 633{~}nm$ (wavelength of
laser), $\zeta=58$ \cite{MAR_97}, $\gamma=0.845{~}dyn\cdot s/cm^2$
, $K_1=0.64 \cdot 10^{-6} {~}dyn,{~}  K_2=0.42 \cdot 10^{-6}
{~}dyn,{~}K_3=0.86 \cdot 10^{-6} {~}dyn$ \cite{Lorenz} ; the
calculations are made for a layer of $50 {~}\mu m$ thickness. For these
parameters $I_c=33.21 {~} W/cm^2$, $\tau=2.49{~}s$. It is easy to show the
following relations: $\gamma_{\perp} \simeq \alpha_on_o/k_0$, 
$\gamma_a \simeq (\alpha_e n_e-
\alpha_o n_o)/k_0$, 
$\varepsilon_\perp \simeq n_o^2$, 
$\varepsilon_a \simeq n_e^2-n_o^2$
(neglecting terms of the order 
$\left(\gamma_{\perp}/(\varepsilon_{\perp}-s_0^2)\right)^2$).

The stability diagram is depicted in Fig. 2a.
The solid line corresponds to a stationary  bifurcation
($Re(\sigma)=Im(\sigma)=0$) and the dashed one corresponds to a 
Hopf bifurcation of the homeotropic state 
($Re(\sigma)=0,Im(\sigma) \neq 0$). These lines divide the 
$(\kappa,\rho)$ plane into a stable and an unstable region of the
homeotropic alignment. They join in a so called Takens-Bogdanov point where
$det({\cal L})=Tr({\cal L})=0$. 

There are two differences compared to the case of a pure LC.
First, the enhancement of the orientational optical nonlinearity
described by the parameter $\zeta$ leads to a "renormalization" of the
threshold intensity (see Eq. (\ref{porog})). (However, since Fig. 2a is
plotted with the renormalized threshold intensity, this does not change the
appearance of the stability diagram). Second, absorption gives rise to the
attenuation of the field inside the nematic. This results in a shift of the
line of primary instability to the region of higher intensities as is shown in
Fig. 2b. From this figure one can see the quantitative difference between the
case when the absorption is neglected (dot-dashed lines) and when the
absorption is taken into account (solid and dashed lines). Note that the
critical intensity $\rho_{th}$ for perpendicular incidence thus becomes larger
than 1.

It must be noted that we supposed that the nematic is maintained at 
constant temperature. Actually, due to the presence of the
absorbing dye, the nematic will be heated by the light \cite{TEMP}. We
have estimated the maximum temperature difference occurring {\it inside} the
nematic 5CB doped with the dye AD1 from the steady-state heat conductivity
equation (one-dimensional since we considered a plane wave). For the range of
intensities $I=30-100 {~}W/cm^2$ this difference was found to be no more than
a few Kelvins. Thus we can usually neglect the temperature dependence of the
material parameters and we took them to be constant across the layer.
\section{Stationary distorted state}
\label{st_dist}
After the homeotropic state looses stability via a stationary bifurcation (at
not too large angle of incidence), the director settles in a stationary
distorted state. One obtains the stationary $z$-dependent reorientation of the
director by variation of the free energy density  (\ref{free_energy}) with
$\bf n$ defined in terms of the angles $\theta,\varphi$. This gives us two
ordinary second-oder differential equations for $\theta(z)$ and $\varphi(z)$
(now these angles can be of arbitrary magnitude). We will not display them
here because they are long and are obtained  straightforwardly. 

These equations contain the field components which obey Maxwell's equations
(\ref{eq_psi}). It is convenient to write (\ref{eq_psi}) using the Oldano
formalism \cite{oldano}.  Following \cite{GAB_00} we write the 
matrix ${\bf\sf D}$ (see (\ref{d})) as  ${\bf\sf D}={\bf\sf D_0}+{\bf\sf
D_z(z)}$, where
\begin{eqnarray}
{\bf\sf D_0}=\left(\matrix{0 & 1-{s_0^2\over e_{\perp}+e_a} & 0 & 0 \cr
e_{\perp} & 0 & 0 & 0 \cr
0 & 0 & 0 & 1 \cr
0 & 0 & e_{\perp} -s_0^2& 0}\right)
\label{D_0}
\end{eqnarray}
and $e_{\perp}=\varepsilon_{\perp}+i\gamma_{\perp}$,
\begin{figure}
\epsfxsize=7.true cm
\makebox[\columnwidth]{\epsffile{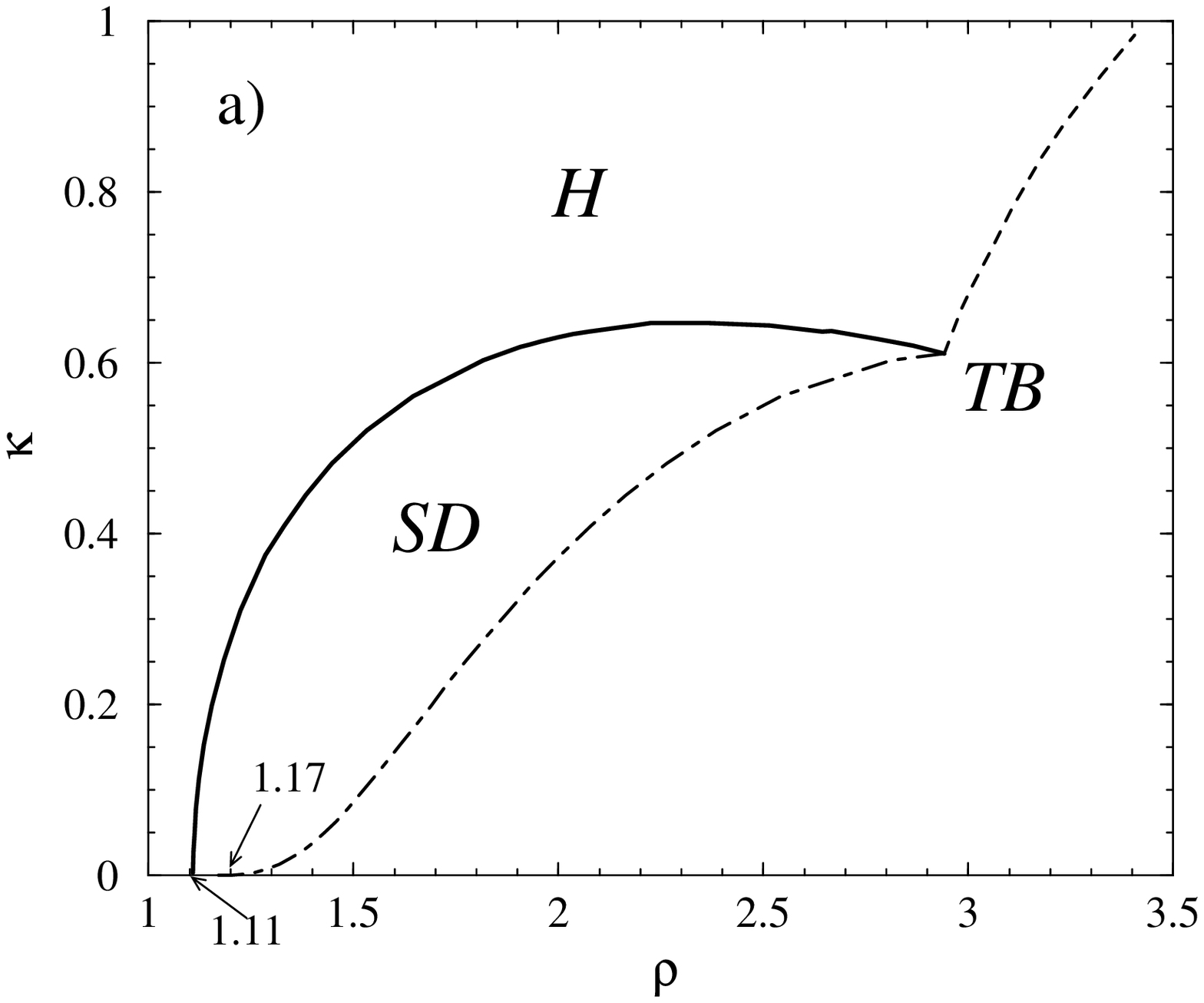}}
\epsfxsize=7.true cm
\makebox[\columnwidth]{\epsffile{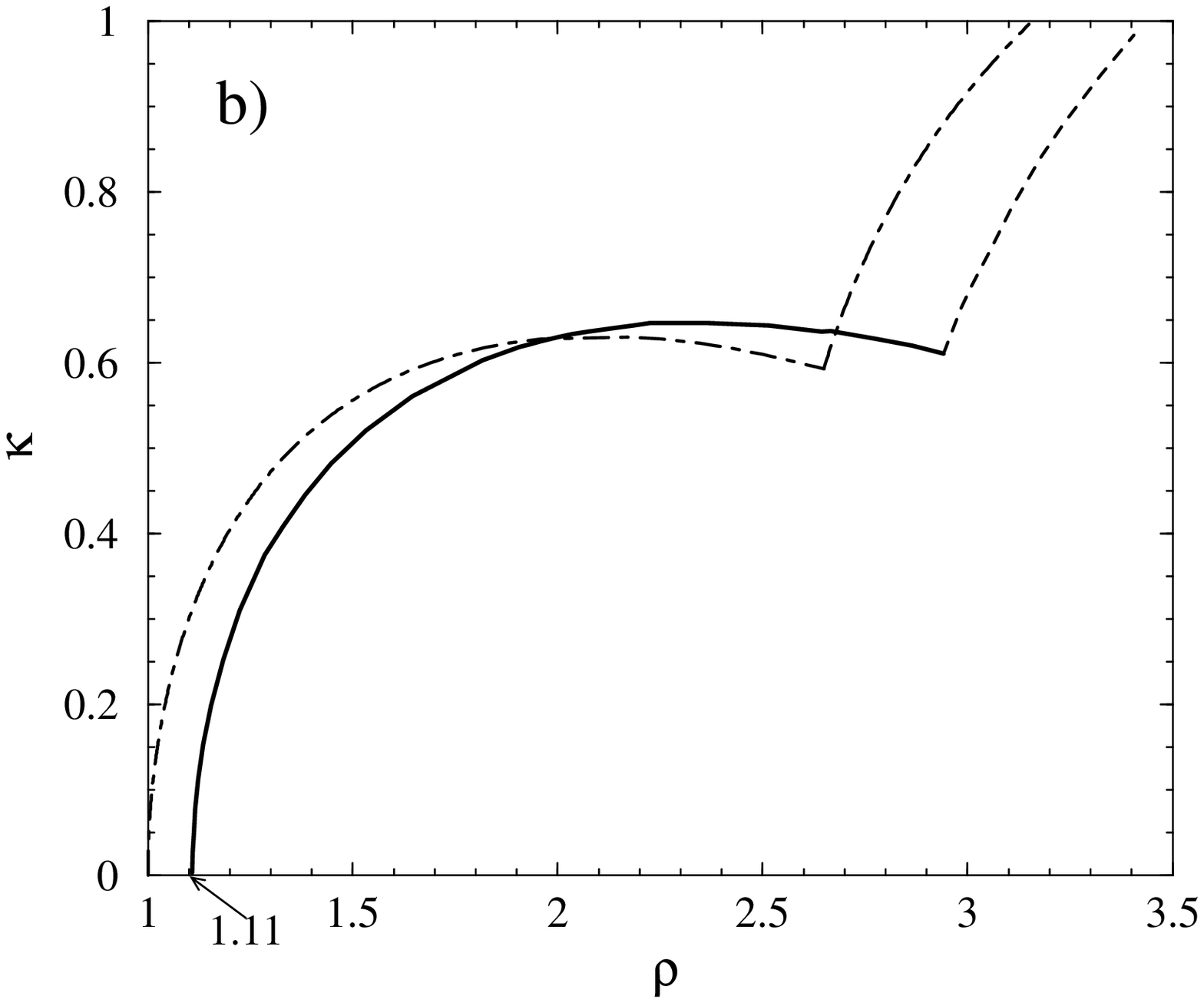}}
\epsfxsize=7.true cm
\makebox[\columnwidth]{\epsffile{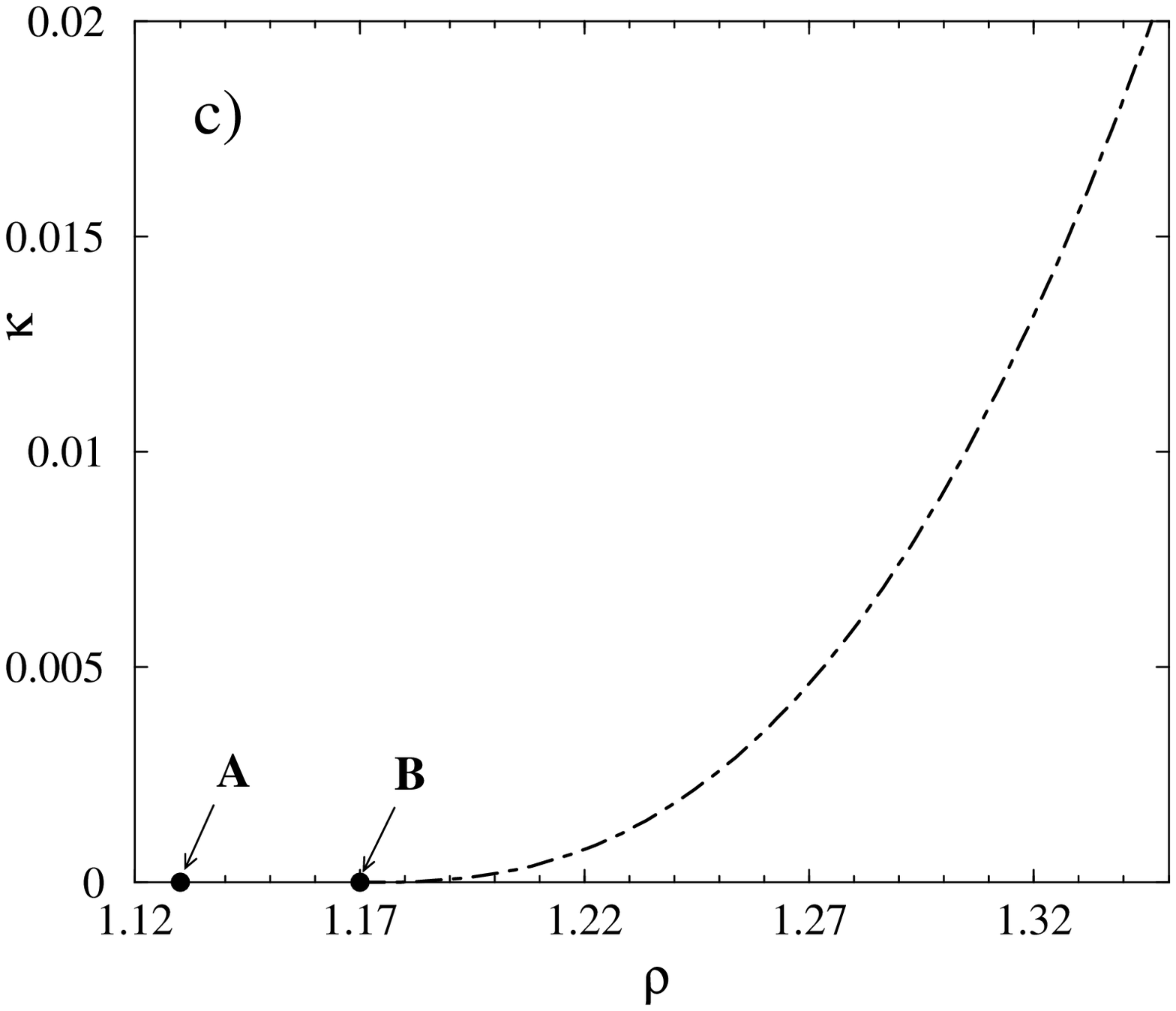}}
\caption{a) Stability diagram of the homeotropic and 
stationary distorted states in the $(\kappa,\rho)$ plane. {\it H}           
is the region of the homeotropic state. {\it SD} is the region of the
stationary distorted state bounded toward large $\rho$ by the secondary Hopf
bifurcation (dash-dotted line). {\it TB} is the Takens-Bogdanov point. 
b) Solid and dashed lines correspond to those in Fig. 2a. The dot-dashed lines
are obtained when the absorption effect is neglected. c) The secondary
instability for small angles of incidence. Points A and B show the
instabilities of the stationary distorted state for perpendicular incidence of
the light with $n_x(z)=0$ and $n_x(z)\neq 0$ correspondingly.}  
\end{figure}
$e_a=\varepsilon_a+i\gamma_a$. The matrix ${\bf\sf D_z}$ contains the
$z$-dependent angles $\theta$ and $\varphi$. It is convenient to introduce a
representation in terms of eigenfunctions of ${\bf\sf D_0}$. The eigenvalue
problem ${\bf\sf D_0} \bar{\alpha}_i=a_i \bar{\alpha}_i$ is solved by the
eigenvalues 
\begin{eqnarray}
&&a_2=-a_1=\sqrt{e_{\perp}-s_0^2},
\nonumber\\
&&a_4=-a_3=\sqrt{\frac{(e_{\perp}+e_a-s_0^2)e_{\perp}}{e_{\perp}+e_a}}
\end{eqnarray}
and eigenvectors 
\begin{equation}
\bar{\alpha}_{1,2}=\left(\matrix{0\cr0\cr
{\mp 1/a_2}\cr 1}\right) 
\mbox{~~,~~}
\bar{\alpha}_{3,4}=\left(\matrix{
{\mp a_4/e_{\perp}}
\cr 1 \cr 0 \cr 0 }\right) 
\label{eigenvecs}
\end{equation}
We introduce the metric tensor
\begin{eqnarray}
{\bf\sf M}=\left(\matrix{ 0&1&0&0 \cr 1&0&0&0 \cr 0&0&0&1 \cr 0&0&1&0}\right) 
\end{eqnarray}
to define a scalar product between these vectors.
With such a metric the eigenvectors are orthogonal to one another:   
$\bar{\alpha}_i^T{\bf\sf M}\bar{\alpha}_j=\delta_{ij}N_i$, where
$N_i$ is the "norm" of vector $\bar{\alpha}_i$. The matrix ${\bf\sf D_0}$ is 
expressed by means of the vectors  $\bar{\alpha}_i$ as 
${\bf\sf D_0}=\sum_i (a_i/N_i)\bar{\alpha}_i\bar{\alpha}_i^T {\bf\sf M}$.
The four vectors $\bar{\alpha}_i$ give the polarization of four
"proper" waves that propagate inside the layer without changing their state of
polarization in the case of homeotropic alignment. The magnitudes of $a_i$
give the indices of refraction of these waves. Two of these vectors
$\bar{\alpha}_1$ ($\bar{\alpha}_2$) correspond to backward (forward)
propagating ordinary waves  and the other two $\bar{\alpha}_3$
($\bar{\alpha}_4$) correspond to backward (forward) propagating extraordinary
waves. The contribution of the backward waves is negligibly small because  the
dielectric properties of the nematic change little on the spatial  scale of
the wavelength \cite{GAB_00}. Thus we can expand $\bar{\Psi}(z)$ as follows
\begin{eqnarray}
\label{psi_definition}
\bar{\Psi}(z)= b_2(z)e^{ik_0a_2z}\bar{\alpha}_2+
b_4(z)e^{ik_0a_4z}\bar{\alpha}_4
\end{eqnarray}
and write (\ref{eq_psi}) in terms
of the amplitudes $b_2(z)$ and $b_4(z)$: 
\begin{eqnarray}
\label{sys_1}
\left\{\begin{array}{ll}
\frac{db_2}{dz}=\frac{ik_0}{N_2}\left[P_{22}(z)b_2+b_4 e^
{-ik_0(a_2-a_4)z}P_{24}(z)\right] \\ 
\frac{db_4}{dz}=\frac{ik_0}{N_4}\left[P_{44}(z)b_4+b_2 e^
{-ik_0(a_4-a_2)z}P_{42}(z)\right], 
\end{array}\right.
\end{eqnarray}
where $P_{kj}(z)=\bar{\alpha}_k^T{\bf\sf M}{\bf\sf D_z}(z)\bar{\alpha_j}$ are the
matrix elements of ${\bf\sf D_z}$ between the eigenvectors:
\begin{widetext}
\begin{eqnarray}
\label{matr_el}
P_{22}&=&\frac{e_a e_{\perp} \cos(\theta)^2 \sin(\varphi)^2}
{a_2^2(e_{\perp}+e_a\cos(\theta)^2\cos(\varphi)^2)},{~}
P_{24}=P_{42}=\frac{\left[a_4 \sin(\theta)-\cos(\theta)\cos(\varphi) s_0 \right]
\sin(\varphi)\cos(\theta) e_a}{a_2 (e_{\perp}+e_a\cos(\theta)^2\cos(\varphi)^2)},
\nonumber\\
P_{44}&=&\frac{e_a \left[ (a_4 \sin(\theta)^2-\sin(2\theta)\cos(\varphi)s_0)
(e_{\perp}+e_a)a_4+e_{\perp} s_0^2(\cos(\theta)^2\cos(\varphi)^2-1)\right]} 
{e_{\perp} (e_{\perp}+e_a) (e_{\perp}+e_a\cos(\theta)^2\cos(\varphi)^2)} 
\end{eqnarray}
\end{widetext}

The advantage of the system (\ref{sys_1}) is that we now have only 
two equations for the "slow" amplitudes $b_2(z)$ and $b_4(z)$. 
So, we have a system of coupled ordinary differential equations for 
$\theta(z)$, $\varphi(z)$, $b_2(z)$ and $b_4(z)$ with boundary conditions 
$\theta |_{z=0,L}=\varphi |_{z=0,L}=0$, and initial conditions
$b_2 |_{z=0}=A_0$, $b_4 |_{z=0}=0$. Here $A_0$ 
can be related to the normalized intensity $\rho$ defined in the
previous section:  
$A_0=\sqrt{8\pi^3\left(\varepsilon_a+\varepsilon_{\perp}\right)
\left(\varepsilon_{\perp}-s_0^2+i \gamma_{\perp}\right)\rho/
(\varepsilon_a\varepsilon_{\perp}\eta})$.  

The system of ``nematic+field'' equations (with boundary conditions) is
invariant under the transformation $\left[\theta,\varphi,E_x, E_y\right]$
$\rightarrow$ $\left[\theta,-\varphi,E_x,-E_y \right]$ owing to the
reflection symmetry with respect to the $y$ direction. Since the primary
instability breaks this symmetry, two different distorted states
exist, which are mutual images under this transformation. For perpendicular
incidence of the light there is an additional reflection symmetry with respect
to the $x$ direction and, as a consequence, the system of
equations is also invariant under the transformation
$\left[\theta,\varphi,E_x, E_y\right]$ $\rightarrow$
$\left[-\theta,\varphi,-E_x,E_y \right]$. 

The system of equations can only be solved numerically.  For this purpose we
introduced the new variables $d\theta/dz,d\varphi/dz$ to transform our set
of equations to a system of six first-oder equations which was solved by the
shooting method. To guarantee that we obtain the solution which originates 
from the homeotropic state we started with intensities only slightly above the
threshold. Then, we increased  $\rho$ slightly and used the values of 
$d\theta/dz |_{z=0},{~}d\varphi/dz |_{z=0}$ obtained in the previous step as
an initial guess.  This procedure allowed us to derive the profiles 
$\theta(z)$, $\varphi(z)$, $b_2(z)$ and $b_4(z)$ for any $\kappa$ and $\rho$
above threshold.

The director and field distributions for $\rho=2.0$ and $\beta_0=11^{\circ}$
($\kappa=0.375$) are shown in Figs. 3,4. In the next section we will show
that for $\kappa=0.375$ the stationary distorted state becomes unstable
at $\rho_c=2.01$, thus these figures represent the state slightly below
the secondary instability. 
\section{Stability analysis of the stationary distorted state}
\label{stab_anal}
Next we have performed a linear stability analysis of the distorted stationary
state with respect to spatially periodic perturbations in the plane of the 
nematic layer. We write  
\begin{eqnarray}
&&{\bf n}={\bf n}_0(z)+\delta {\bf n} (x,y,z,t)=
\nonumber\\
&&{\bf n}_0(z)+\delta {\bf n}(z) e^{\sigma t +i (q x+p y)},{~~~~}
\bar{\Psi}=\bar{\Psi}_0+\bar{\Psi}_1=
\nonumber\\
&&\sum_{k=2,4} (b_k(z)+\delta b_k(z) 
e^{\sigma t +i (q x+p y)})
e^{ik_0a_k z}\bar{\alpha}_k, 
\label{n_space}
\end{eqnarray}
where $\delta {\bf n}$ and $\delta b_k$ are small spatially periodic
perturbation with wavenumbers $q$ and $p$; $\sigma$ is the growth
rate. 

From the equation ${\bf n}^2=1$ follows that $\bf n_0 \delta \bf n=0$. Thus
there are only two independent components of $\delta\bf n$. We obtained two
linear equations for $\delta n_x(z)$ and $\delta n_y(z)$ which contain
$\delta n_x(z)$, $\delta n_y(z)$ itself, their $z$ derivatives up to  second
oder and $\delta b_{2,4}(z)$ with complicated coefficients depending on the
stationary distorted state ${\bf n}_0(z),b_{2,4}(z)$. Also, we decomposed the
matrix ${\bf\sf D}$ (see (\ref{d})) as ${\bf\sf D}={\bf\sf D_0}+{\bf\sf
D_z(z)}+{\bf\sf D_1(\delta \bf n)}$, where the matrices ${\bf\sf D_0},{\bf\sf
D_z(z)}$  correspond to the stationary state and were defined in the previous
section, and the matrix ${\bf\sf D_1(\delta \bf n)}$ depends linearly on
$\delta \bf n$. After linearization of (\ref{eq_psi}) the equations for
$\delta b_{2,4}$ can be obtained: 
\begin{widetext}
\begin{eqnarray}
\label{sys_2}
\left\{\begin{array}{ll}
\frac{d(\delta b_2(z))}{dz}=\frac{ik_0}{N_2}\left( \delta b_2 P_{22}
+\delta b_4  e^{ik_0(a_4-a_2)z}P_{24}+
b_2 P_{22}^{(1)} +b_4 e^{ik_0(a_4-a_2)z}
P_{24}^{(1)}\right)\\ 
\frac{d(\delta b_4(z))}{dz}=\frac{ik_0}{N_4}\left( \delta b_2               
e^{ik_0(a_2-a_4)z} P_{24} +\delta b_4 P_{44} +
b_2 e^{ik_0(a_2-a_4)z} P_{24}^{(1)} +b_4 
P_{44}^{(1)}\right),  \end{array}\right. 
\end{eqnarray}
\end{widetext}
where $P^{(1)}_{kj}=\bar{\alpha}_k^T{\bf\sf M}{\bf\sf D_1}\bar{\alpha_j}$ are
the matrix elements of ${\bf\sf D_1}$ with respect to the eigenvectors
(\ref{eigenvecs}) and the $P_{kj}$ were defined in (\ref{matr_el}).
 
We have linearized the equation (\ref{eq_psi}) substituting $\psi$ in the form 
(\ref{n_space}). In principle we should have started
from  Maxwell's equations because the field perturbations contain $x,y$
dependence. However this is a very good approximation because the
corrections are of the oder $q/k_0$, $p/k_0 <<1$. 

To solve the eigenvalue problem for $\sigma$ we expand  $\delta n_{x,y}(z)$,
$\delta b_{2,4}(z)$ with respect to $z$ in systems of functions which satisfy
the boundary conditions (Galerkin method).
\begin{figure}
\epsfxsize=7.true cm
\makebox[\columnwidth]{\epsffile{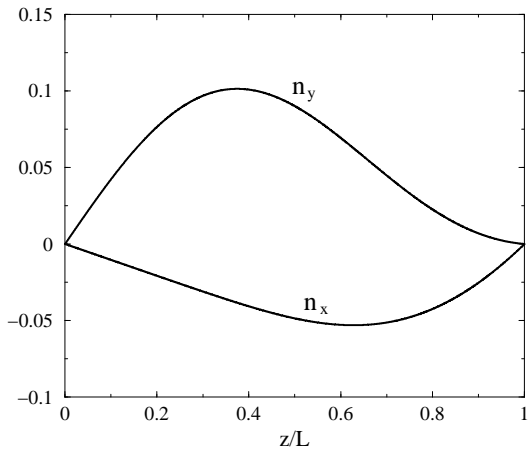}}
\caption{ Profiles of the director components $n_x$, $n_y$ for the stationary
distorted state at $\rho=2.0$ and $\beta_0=11^{\circ}$ ($\kappa=0.375$).} 
\end{figure}
\begin{figure}
\epsfxsize=7.true cm
\makebox[\columnwidth]{\epsffile{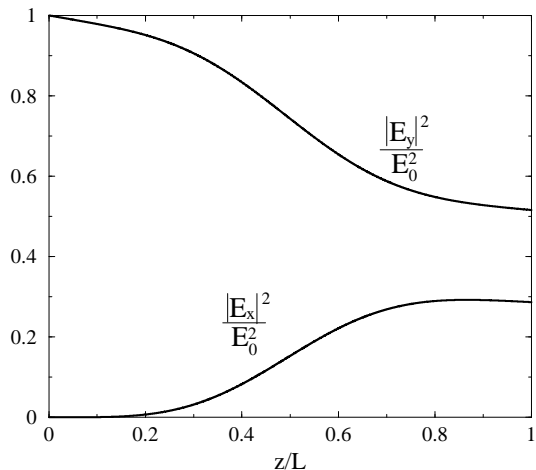}}
\caption{Distortion of the field components inside the nematic layer for the
stationary  distorted state ($\rho=2.0$, $\beta_0=11^{\circ}$). $E_z$ is small
compared to $E_x$, $E_y$ and is not depicted; $E_0$ is the amplitude of the
incident electric field.}
\end{figure}
\begin{figure}
\epsfxsize=7.true cm
\makebox[\columnwidth]{\epsffile{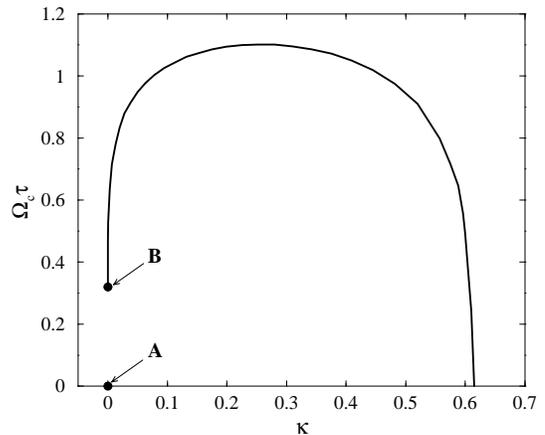}}
\caption{ Dimensionless Hopf frequency $\Omega_c \tau$ for the secondary 
instability versus $\kappa$. Points A and B are the Hopf frequencies  at
points A and B depicted in Fig. 2 c).} 
\end{figure}
For $\delta {\bf n}$ the boundary conditions are $\delta
n_{x,y}|_{z=0,L}=0$, thus we write  $\delta {\bf n}$=$\sum_{k} {\bf A}_k
\sin(\pi k z/L)$.  Clearly the boundary
conditions for the perturbations of the field amplitudes are $\delta
b_{2,4}|_{z=0}=0$. 
One can see that at  $z=0,L$ the r.h.s. of the system
(\ref{sys_2}) vanishes so one also has  $d(\delta b_{2,4})/dz|_{z=0,L}=0$.
Therefore we used the expansion $\delta {\bf b}=\sum_{n} {\bf B}_n \sin^2(\pi
n z/(2L))$. This set of functions is complete but not orthogonal. We have to
truncate these  expansions to a finite number of modes.  

We have solved the eigenvalue problem numerically to find the
neutral surface $\rho_0(q,p)$ (for given angle $\beta_0$) which is defined by
the condition $Re(\sigma(q,p))=0$. The number of Galerkin modes was chosen 
such that the accuracy of the calculated eigenvalues was better than $0.1\%$
(we took six modes for $\delta {\bf n}$ and forty modes for            
$\delta {\bf b}$). The minimum of this surface gives the critical intensity 
$\rho_c$=$\min_{q,p}\rho_0(q,p)$ and the critical wavevector $(q_c,p_c)$.
Since $\Omega_c=Im(\sigma)$ turned out to be nonzero at the minimum, the
instability corresponds to a Hopf bifurcation. The branch of the secondary
Hopf instability is depicted as the dash-dotted line in Fig. 2a and for small
angles of incidence in Fig. 2c.  It is interesting to note the  following
tendencies: as the incident angle $\beta_0$ increases the critical intensity
also increases, but the director  and field  deformations at the secondary
instability decrease. 

The dimensionless Hopf frequency $\Omega_c \tau$ ($\tau$ is defined in
section \ref{section_lin}) versus $\kappa$ is shown in
Fig. 5.  Figure 6 shows a typical contour plot of the neutral surface
$\rho_0(q,p)$. The point $(q_c L,p_c L)$ in this figure is the minimum of the
surface and as is seen the  bifurcation is inhomogeneous with some critical
vector $(q_c,p_c) \ne {\vec 0}$. This means that travelling waves are
expected to appear. $\rho_c$ is only slightly below the homogeneous
threshold $\rho_0(q=0,p=0)$, which was calculated before
for the pure LC \cite{GAB_99,GAB_00}.    

As was pointed out in section \ref{st_dist}, for nonzero $\beta_0$ there are
two symmetry-degenerate stationary distorted states. Clearly the two neutral
surfaces are related by changing $p$ to $-p$ and the critical wave vectors will
be $(q_c,p_c)$ and $(q_c,-p_c)$. Thus two different travelling waves with
critical vectors $(q_c,\pm p_c)$ can be realized depending on which stationary
state will be selected after the homeotropic state loses stability. 

An interesting situation arises in the limit of normal incidence. One might
expect that for $\beta_0 \rightarrow 0$ the wavenumber $q_c \rightarrow 0$,
since in this limit the external symmetry breaking in the $x$ direction
vanishes. However, this turned out not to be the case.
The reason is that then another stationary instability that spontaneously
breaks $x$-reflection symmetry intervenes the primary and the Hopf
bifurcation. For  the parameters of our computation one has $\rho_{th}=1.11$,
$\rho_{c1}=1.13$ (point A in Fig. 2c) and $\rho_{c2}=1.17$ (point B in Fig.
2c). One now has four symmetry-degenerate states and consequently four
traveling waves with critical wave vectors $(\pm q_c, \pm p_c)$.
\begin{figure}
\epsfxsize=7.true cm
\makebox[\columnwidth]{\epsffile{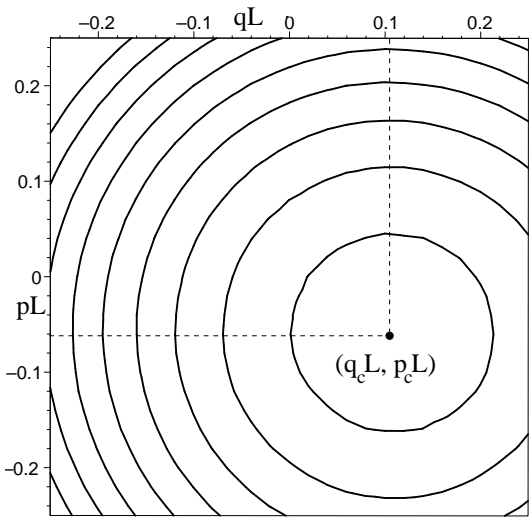}}
\caption{   Contour lines for the surface $\rho(p,q)$ correspond to
$\beta_0=11^{\circ}$ ($\kappa=0.375$). 
The critical intensity is $\rho_c=2.01$ with the critical wavevector $(q_c
L,p_c L)$ = $( 0.11,-0.06)$; $\rho_0(q=0,p=0) -\rho_c$ = $1.5 \cdot
10^{-3}$.}
\end{figure}
In some further investigations we have changed the ratios between the
elastic constants keeping other material parameters constant and saw the
following tendency: the larger the anisotropy of the constants, the deeper
the minimum of the surface becomes and the larger the magnitudes of the
critical wavenumbers (see Fig. 7a,b). The absolute error of the dimensionless
critical wavenumbers $q_c L,p_c L$ depicted in this figure is less than
$10^{-2}$. In the one-constant approximation the bifurcation is homogeneous
($q_c,p_c=0$) for any $\kappa$. This latter can be easily proved analytically.
Perturbation theory can be used to investigate $q,p$ dependence of the
critical eigenvalue of an arbitrary stationary state. The calculations shows,
that the perturbation of the eigenvalue is $\sim p^2+q^2$.

From  Fig. 7a,b one can see that $q_c L, p_c L \sim 0.1$. This means that
the period of the structure $2 \pi/q_c,~2 \pi/p_c \sim 60 L = 0.3 {~}cm$.
Thus in an experiment the spot size of the light must be rather large in
oder to observe the travelling waves.
\begin{figure}[htb]
\epsfxsize=6.8true cm
\makebox[\columnwidth]{\epsffile{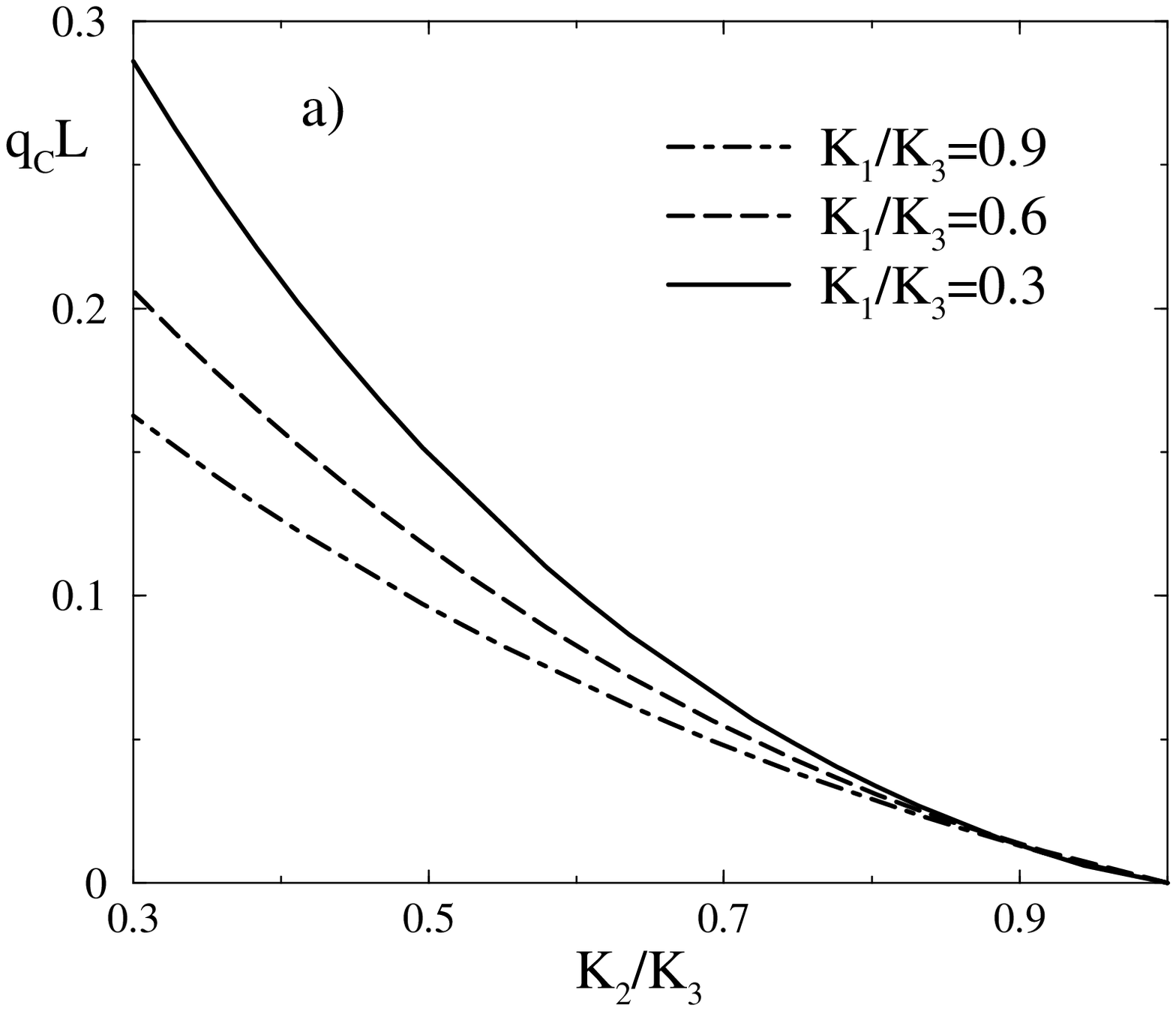}}
\epsfxsize=7.true cm
\makebox[\columnwidth]{\epsffile{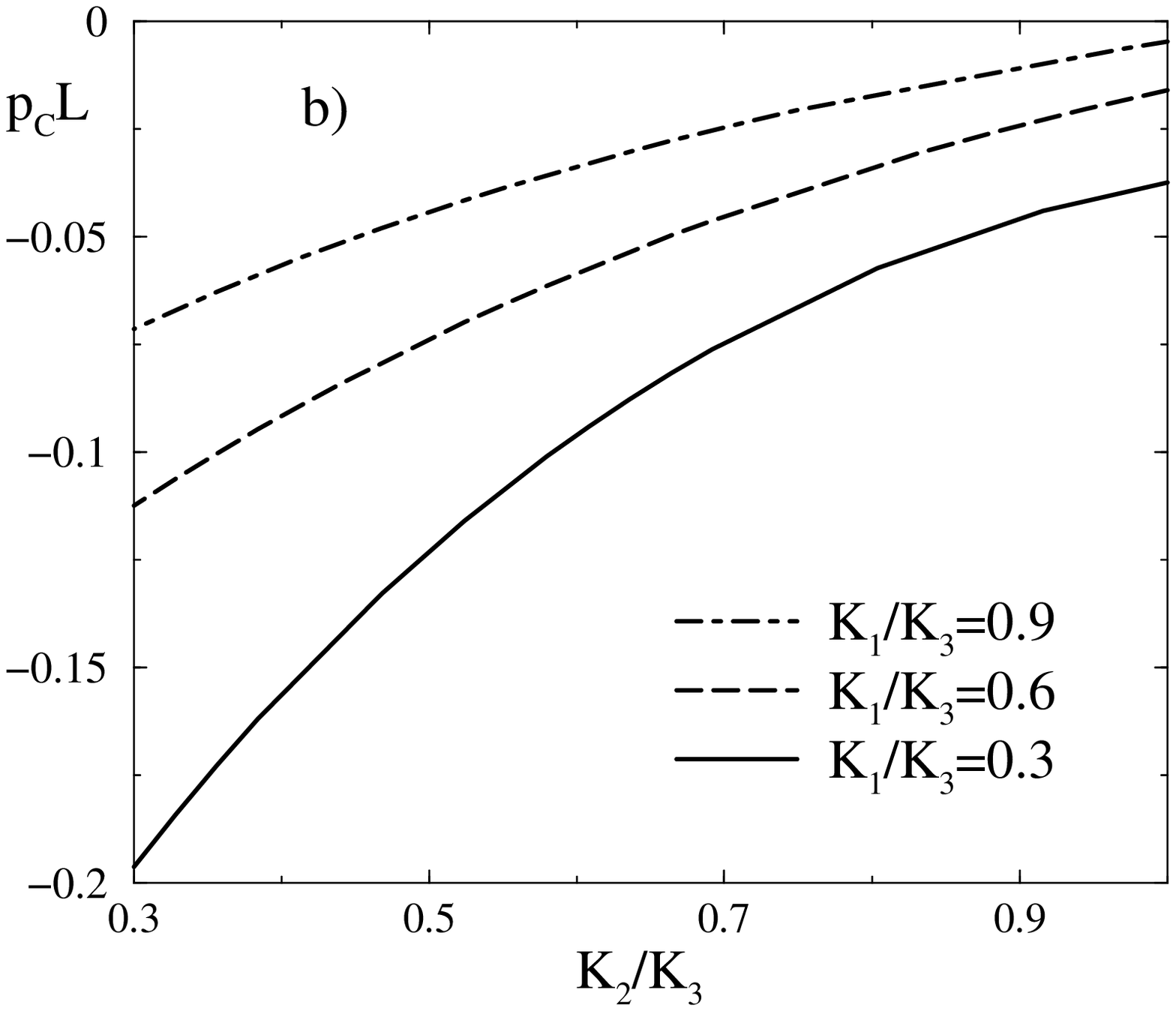}}
\caption{ Critical wavenumbers $q_c , p_c$ versus $K_2/K_3$ for different
ratios $K_1/K_3$ ($\beta_0=11^{\circ}$).}  
\end{figure}

Finally we remark on the behaviour of the system in the nonlinear
regime above the Hopf bifurcation. This system without transverse degrees of
freedom has been studied extensively, and various regimes of complex behaviour
have been discovered. The bifurcation studied in this work marks the
transition to simple periodic oscillations in the system without transverse
degrees of freedom, which is the first step towards complex behaviour.
In models \cite{GAB_00,GAB_99} and simulations \cite{Gab3}, a gluing
bifurcation was found above the secondary Hopf instability, which is a
homoclinic bifurcation that restores the symmetry broken by the Freedericksz
transition. This gluing bifurcation was recently observed experimentaly
\cite{Cipparrone}. After this first gluing, complex nonlinear
behaviour and eventually chaos was observed in both theory, simulation and
experiment \cite{it1}. An analogous gluing bifurcation should exist also in the
case of the spatially extended system. 

The behaviour of the system in the vicinity of this gluing bifurcation, can,
however be radically different from what was observed in the experiment
\cite{Cipparrone}. In the spatially constrained system (i.e. the director
oscillation induced by a narrow beam as observed in the experiments) one
observes stochastic behaviour in the vicinity of the first gluing only as a
consequence of experimental noise. It has been shown, however, \cite{Coul}
that any spatially extended system, which possesses a homogeneous limit cycle
(which is stable with respect to homogeneous perturbations) becomes unstable
as it approaches a homoclinic bifurcation. This instability is either a phase
instability, or a finite-wavelength period-doubling instability. On these
grounds one can expect to observe very complicated behaviour (probably
spatio-temporal chaos) in our system already at the threshold intensity of the
first gluing. As opposed to the previous case, this would be
true deterministic chaos, not merely stochasticity due to noise. 
\section{Conclusion}
We have found the threshold of the LIFT for the homeotropic state and the
threshold of the secondary instability of the stationary distorted state in a
nematic LC, including the dye-doped case, for different incident angles of the
light.  In particular we have demonstrated that the stationary distorted state
loses stability in an inhomogeneous Hopf bifurcation with some nonzero critical
wavenumber that leads to the formation of travelling waves in the plane of
the layer. 

Our result demonstrates a general feature of Hopf bifurcations in spatially
extended systems with broken reflection symmetry, as is the case in the
LIFT-distorted state. Except for special cases, like those where the
reflection symmetry can be restored by going into a moving frame, the neutral
surface exhibits the signature of  the broken symmetry. Consequently, at
$q_c=p_c=0$ the neutral surface does not have a stationary point and thus
cannot have a minimum. This general feature was apparently first noted in the
context of reaction-diffusion systems \cite{Menz}.
\section{Acknowledgments}
We wish to thank A.P. Krekhov for helpful discussions.
This work was supported by DFG grant $N^{\circ}$ KR690/16-1.

\begin{thebibliography}{10.}
%
\bibitem{Tab85} N. V. Tabiryan, A. V. Sukhov and B. Ya. Zel'dovich,
Mol. Cryst. Liquid Cryst. {\bf 136}, 
1-140 (1985).

\bibitem{simoni} F. Simoni, {\em Nonlinear optical properties of liquid 
crystals} (World Scientific, Singapore, 1997).

\bibitem{mag_85} F. Lonberg and R. B. Meyer, Phys. Rev. Lett. {\bf 55}, 718
(1985).

\bibitem{mag_87} D. W. Allender, R. M. Hornreich and D. L. Johnson, 
Phys. Rev. Lett. {\bf 59}, 2654 (1987).

\bibitem{JAN_99} Istvan Janossy, J. Nonlin. Opt. Phys. Mat. 
{\bf 8}, 361 (1999).

\bibitem{MAR_97} L. Marrucci, D. Paparo, P. Maddalena, 
E. Massera, E. Prudnikova, and E. Santamato, J. Chem. Phys 
{\bf 107}, 9783 (1997).

\bibitem{Gen} P. G. de Gennes and J. Prost, {\em The physics of liquid
crystals} (Clarendon press, Oxford, 1993).

\bibitem{Tab_98} N.V. Tabiryan, A.L. Tabiryan-Murazyan, 
V. Carbone, G. Cipparrone, C. Umeton, 
C. Versace, T. Tschudi, Optics Comm. {\bf 154}, 70 (1998).

\bibitem{Lorenz} A. Buka and L. Kramer, {\em Pattern formation in liquid
crystals} (Springer, 1995).

\bibitem{TEMP} I. Janossy and T. Kosa, Mol. Cryst. Liq. Cryst. 
{\bf 207}, 189 (1991).

\bibitem{oldano} C. Oldano, Phys. Rev. A {\bf 40}, 6014 (1989).

\bibitem{GAB_00} G. Demeter, Phys. Rev. E {\bf 61}, 6678 (2000). 

\bibitem{GAB_99} G. Demeter and L. Kramer, Phys. Rev. Lett.
{\bf 83}, 4744 (1999). 

\bibitem{Gab3} G. Demeter and L. Kramer, Phys. Rev. E  {\bf 64}, 020701(R)
(2001).

\bibitem{Cipparrone} V. Carbone, G. Cipparrone, and G. Russo, Phys. Rev. E 
{\bf 63}, 051701 (2001).

\bibitem{it1} G. Cipparrone, V. Carbone, C. Versace, C. Umeton,
R. Bartolino and F. Simoni, Phys. Rev. E {\bf 47}, 3741 (1993).

\bibitem{Coul} M. Argentina, P. Coullet and E. Risler, 
Phys. Rev. Lett. {\bf 86}, 807 (2001); P. Coullet, E. Risler, N.
Vandenberghe, J. Stat. Phys. {\bf 101}, 521 (2000). 

\bibitem{Menz} A. B. Rovinsky and M. Menzinger, Phys. Rev. Lett. {\bf 69},
1193 (1992); {\bf 70}, 778 (1993).

\end{thebibliography}
\end{document}